%% file: ms.tex
\documentclass[letter]{article}
\RequirePackage[log]{snapshot}
\usepackage{iclr2018_conference,times}

\usepackage{pgfplots}
\usepackage{booktabs}   
\usepackage{subcaption} 
\usepackage{wrapfig}
\usepackage{courier}
\usepackage{tikz}
\usetikzlibrary{arrows}

\usepackage{times}
\usepackage{courier}
\usepackage{url}

\usepackage{amssymb}
\usepackage{xspace}
\usepackage{natbib}
\usepackage{amsfonts}
\usepackage{graphicx}
\usepackage{comment}
\usepackage{floatflt}
\usepackage{amsmath}
\usepackage{multicol}
\usepackage{multirow}
\usepackage{enumitem}
\usepackage{color}
\usepackage{listings}
\usepackage{xspace}

\usepackage{algorithm}
\usepackage{algpseudocode}
\algnewcommand\algorithmicinput{\textbf{Input:}}
\algnewcommand\INPUT{\item[\algorithmicinput]}
\algnewcommand\algorithmicoutput{\textbf{Output:}}
\algnewcommand\OUTPUT{\item[\algorithmicoutput]}

\newtheorem{problem}{Problem}
\DeclareMathOperator*{\argmax}{arg\,max}

\newcommand{\system}{\textsc{Bayou}\xspace}

\renewcommand{\paragraph}[1]{\smallskip\noindent {\bf #1~~}}

\newcommand{\GSNN}{{\sc Gsnn}\xspace}

\newcommand{\secref}[1]{Section~\ref{sec:#1}}
\newcommand{\seclabel}[1]{\label{sec:#1}}

\newcommand{\figref}[1]{Figure~\ref{fig:#1}}
\newcommand{\figlabel}[1]{\label{fig:#1}}

\newcommand{\whilec}{\textbf{while}}
\newcommand{\skipc}{\textbf{skip}}
\newcommand{\ifc}{\textbf{if}}
\newcommand{\thenc}{\textbf{then}}
\newcommand{\elsec}{\textbf{else}}
\newcommand{\doc}{\textbf{do}}
\newcommand{\tryc}{\textbf{try}}
\newcommand{\catchc}{\textbf{catch}}
\newcommand{\declc}{\textbf{let}}
\newcommand{\letc}{\textbf{let}}
\newcommand{\callc}{\textbf{call}}

\newcommand{\id}[1]{\mathsf{#1}}

\newcommand{\lang}{\textsc{Aml}\xspace}
\newcommand{\Prog}{\mathit{Prog}}
\newcommand{\Sprog}{\mathsf{Prog}}

\newcommand{\Sy}{\mathsf{Y}}
\newcommand{\Sx}{\mathsf{X}}
\newcommand{\Sz}{\mathsf{Z}}
\newcommand{\sY}{\mathsf{Y}}
\newcommand{\sX}{\mathsf{X}}
\newcommand{\sProg}{\mathsf{Prog}}
\newcommand{\sZ}{\mathsf{Z}}
\newcommand{\M}{\theta}
\newcommand{\D}{\mathcal{D}}
\newcommand{\Mstar}{\mathbf{M^*}}

\newcommand{\Expec}{\mathbf{E}}

\newcommand{\I}{\mathbf{I}}

\newcommand{\old}[1]{}

\newcommand{\Senv}{\mathsf{Env}}
\newcommand{\Env}{\mathit{Env}}
\newcommand{\safe}{\mathit{safe}}

\newcommand{\Sh}{\mathsf{H}}

\newcommand{\Next}{\mathit{Next}}

\newcommand{\Kws}{\mathit{Keys}}
\newcommand{\Types}{\mathit{Types}}
\newcommand{\Calls}{\mathit{Calls}}

\renewcommand{\bar}{\overline}

\newcommand{\Eqv}{\mathit{Eqv}}

\makeatletter
\DeclareRobustCommand*\cal{\@fontswitch\relax\mathcal}
\makeatother

\newcommand{\modelname}{\textsc{Ged}\xspace}

\title{Neural Sketch Learning for Conditional \\ Program Generation}
\author{Vijayaraghavan Murali, Letao Qi, Swarat Chaudhuri,  and 
Chris Jermaine\\
Department of Computer Science \\
Rice University\\
Houston, TX 77005, USA.\\
$\mathtt{\{vijay, letao.qi, swarat, cmj4\}@rice.edu}$
}
\iclrfinalcopy

\begin{document}
\maketitle

\begin{abstract}
\input{abstract}
\end{abstract}

\section{Introduction}\label{sec:intro}
\input{intro}

\section{Problem Statement}\seclabel{overview}

\input{overview}



\section{Technical Approach}\seclabel{approach}

\input{bayesian}


\section{Learning}\seclabel{learning}

\input{variational}

\section{Combinatorial Concretization}\seclabel{synthesis}
\input{synthesis-short}


\section{Experiments}\seclabel{eval}

\input{eval}

\section{Related Work}\seclabel{relwork}

\input{relwork}

\section{Conclusion}\seclabel{conc}

\input{conc}

\paragraph{Acknowledgements} This research was supported by DARPA MUSE
award \#FA8750-14-2-0270 and a Google Research Award. 

{\small 
\bibliography{refs,refs-2,refs-3,refs4}
}
\newpage 
\appendix


\input{overview-appendix}

\section{The \lang Language}\label{aml-appendix}

\input{aml-appendix}

\section{Abstracting \lang Programs into Sketches}\label{abs-appendix}

\input{abs-appendix}

\section{Neural Network Details}
\label{appendix-neural}

\input{encoder}

\input{neural}



\section{Additional Evaluation}

\input{eval-appendix}

\end{document}

%% file: abstract.tex
We study the problem of generating source code in a strongly typed,
Java-like programming language, given a label (for example a set of
API calls or types) carrying a small amount of information about the
code that is desired. The generated programs are expected to respect a
``realistic'' relationship between programs and labels, as exemplified
by a corpus of labeled programs available during training.

Two challenges in such {\em conditional program generation} are that
the generated programs must satisfy a rich set of syntactic and
semantic constraints, and that source code contains many low-level
features that impede learning.  We address these problems by training
a neural generator not on code but on \emph{program sketches}, or
models of program syntax that abstract out names and operations that
do not generalize across programs. During generation, we infer a
posterior distribution over sketches, then concretize samples from
this distribution into type-safe programs using combinatorial
techniques.  We implement our ideas in a system for generating
API-heavy Java code, and show that it can often predict the entire
body of a method given just a few API calls or data types that appear
in the method.

%% file: intro.tex
Neural networks have been successfully applied to many generative
modeling tasks in the recent
past~\citep{oord2016pixel,ha2017neural,vinyals2015show}. However, 
the use of these models in generating highly structured text remains relatively
understudied. In this paper, we present a method, combining neural and
combinatorial techniques, for the condition generation of an important
category of such text: {\em the source code of programs in
Java-like programming languages}.

The specific problem we consider is one of supervised learning. During
training, we are given a set of programs, each program annotated with a 
label, which may contain information such as the set of 
API calls or the types used in the code.  
Our goal is to learn
a function $g$ such that for a test case of the form $(\sX, \sProg)$ (where
$\sProg$ is a program and $\sX$ is a label), $g(\sX)$ is a compilable,
type-safe program that is equivalent to $\sProg$. 

This problem has immediate applications in helping humans 
solve programming
tasks~\citep{hindle2012naturalness,raychev2014code}. In the usage
scenario that we envision, a human programmer uses a label to specify
a small amount of information about a program that they have in mind. Based on this
information, our generator seeks to produce a program equivalent to
the ``target'' program, thus performing a particularly powerful
form of code completion.

Conditional program generation is a special case of {\em program
  synthesis}~\citep{mannawaldinger, summers1977methodology}, the
  classic problem of generating a program given a constraint on its
  behavior. This problem has received significant interest in recent
  years ~\citep{sygus,gulwani-survey}. In particular, several neural
  approaches to program synthesis driven by {\em input-output
  examples} have
  emerged~\citep{balog2016deepcoder,parisotto2016neuro,robustfill}.
  Fundamentally, these approaches are tasked with associating a
  program's syntax with its semantics. As doing so in general is
  extremely hard, these methods choose to only generate programs in
  highly controlled domain-specific languages. For example,
  \citet{balog2016deepcoder} consider a functional language in which
  the only data types permitted are integers and integer arrays,
  control flow is linear, and there is a sum total of 15 library
  functions. Given a set of input-output examples, their method
  predicts a vector of binary attributes indicating the presence or
  absence of various tokens (library functions) in the target program,
  and uses this prediction to guide a combinatorial search for
  programs.

In contrast, in conditional program generation, we are already given a
set of tokens (for example library functions or types) that appear in a
program or its metadata. Thus, we sidestep the problem of learning the semantics of
the programming language from data. We ask: does this simpler setting
permit the generation of programs from a much richer, Java-like
language, with one has thousands of data types and API methods,
rich control flow and exception handling, and a strong type system?

While simpler than general program synthesis, this problem is still
highly nontrivial. Perhaps the central issue is that to be acceptable
to a compiler, a generated program must satisfy a rich set of
structural and semantic constraints such as ``do not use undeclared
variables as arguments to a procedure call'' or ``only use API calls
and variables in a type-safe way''. Learning such constraints
automatically from data is hard. Moreover, as this is also a
supervised learning problem, the generated programs also have to
follow the patterns in the data while satisfying these constraints.

We approach this problem with a combination of neural learning and
type-guided combinatorial search~\citep{example-pldi15}. Our central
idea is to learn not over source code, but over
tree-structured syntactic models, or {\em sketches}, of programs. A
sketch abstracts out low-level names and operations from a program,
but retains information about the program's control structure, the
orders in which it invokes API methods, and the types of arguments and
return values of these methods. We propose a particular kind of probabilistic
encoder-decoder, called a Gaussian Encoder-Decoder or \modelname, 
to learn a distribution over sketches conditioned on labels. During
synthesis, we sample sketches from this distribution, then flesh out
these samples into type-safe programs using a 
combinatorial method for program synthesis. Doing so effectively is possible
because our sketches are designed to contain rich information about
control flow and types.

We have implemented our approach in a system called
\system.\footnote{\system is publicly available at \url{https://github.com/capergroup/bayou}.} We
evaluate \system in the generation of API-manipulating Android
methods, using a corpus of about 150,000 methods drawn from
an online repository.  Our experiments show that \system can often
generate complex method bodies, including methods implementing tasks
not encountered during training, given a few tokens as input.

%% file: overview.tex
Now we define {conditional program generation}.  Assume a universe $\mathbb{P}$ of {\em programs} and a universe $\mathbb{X}$ of {\em
  labels}. Also 
assume a set of training examples of the form
$\{ (\sX_1, \sProg_1), (\sX_2, \sProg_2), ... \}$, where each $\sX_i$
is a label and each $\sProg_i$ is a program. These examples 
are sampled from an unknown distribution $Q(X, \Prog)$,
where $X$ and $\Prog$ range over labels and 
programs, respectively.\footnote{We use italic fonts for
  random variables and sans serif --- for example $\Sx$ --- for
  values of these variables.} 

We assume an equivalence relation $\Eqv \subseteq \mathbb{P} \times \mathbb{P}$ over
programs. If $(\sProg_1, \sProg_2) \in \Eqv$,
then $\sProg_1$ and $\sProg_2$ are \emph{functionally
  equivalent}.  The definition of functional equivalence differs
across applications, but in general it asserts that two
programs are ``just as good as'' one another.  


The goal of {\em conditional program generation} is to use the
training set to learn a function
$g : \mathbb{X} \rightarrow \mathbb{P}$ such that the expected value
$\Expec [I((g(X), \Prog) \in \Eqv)]$ is maximized.  Here, $I$ is the
indicator function, returning 1 if its boolean argument is true, and 0
otherwise.  Informally, we are attempting to learn a function $g$ such
that if we sample $(\sX, \sProg) \sim Q(X, Prog)$, $g$ should be able
to reconstitute a program that is functionally equivalent to $\sProg$,
using only the label $\sX$.

\subsection{Instantiation}  

In this paper, we consider a
particular form of conditional program generation.  We take the domain
$\mathbb{P}$ to be the set of possible programs in a 
programming language called \lang that captures the essence of
API-heavy Java programs (see
Appendix~\ref{aml-appendix} for more details).  \lang includes complex control flow
such as loops, \texttt{if}-\texttt{then} statements, and exceptions;
access to Java API data types; and calls to Java API methods. \lang is
a strongly typed language, and by definition, $\mathbb{P}$ only
includes programs that are type-safe.\footnote{In research on
  programming languages, a program is typically judged as type-safe
  under a {\em type environment}, which sets up types for the
  program's input variables and return value. Here, we consider a
  program to be type-safe if it can be typed under {\em some} type
  environment.}
To define labels, we assume three finite sets: a set $\Calls$ of
possible API calls in \lang, a set $\Types$ of possible object types,
and a set $\Kws$ of {\em keywords}, defined as words, such as ``read''
and ``file'', that often appear in textual
descriptions of what
programs do. The
space of possible labels is $\mathbb{X} = 2^{Calls} \times 2^{Types}
\times 2^{\Kws}$
(here $2^S$ is the power set of $S$).

Defining $\Eqv$ in practice is tricky. For example, a reasonable definition
of $\Eqv$ is that $(\sProg_1, \sProg_2) \in Eqv$ iff $\sProg_1$ and
$\sProg_2$ produce the same outputs on all inputs.  But given the
richness of \lang, the problem of
determining whether two \lang programs always produce the same
output is undecidable.  As such, in practice we can only measure
success indirectly, by checking whether the programs use the same
control structures, and whether they can produce the same API call
sequences.  We will discuss this issue more in \secref{eval}.

\begin{figure}[t]
\begin{tabular}{cc}
\begin{minipage}{0.5\textwidth}
\begin{lstlisting}[language=Java, basicstyle=\ttfamily\scriptsize]
  String s;
  BufferedReader br;
  FileReader fr;
  try {
   fr = new FileReader($String);
   br = new BufferedReader(fr);
   while ((s = br.readLine()) != null) {}
   br.close();
  } catch (FileNotFoundException _e) {
  } catch (IOException _e) {
 }
\end{lstlisting}
\end{minipage}
&
\begin{minipage}{0.5\textwidth}
\begin{lstlisting}[language=Java, basicstyle=\ttfamily\scriptsize]
 String s;
 BufferedReader br;
 InputStreamReader isr;
 try {
  isr = new InputStreamReader($InputStream);
  br = new BufferedReader(isr);
  while ((s = br.readLine()) != null) {}
 } catch (IOException _e) {
 }
\end{lstlisting}
\end{minipage}
\\ (a) & (b) \\
\end{tabular}
\hfill
\caption{Programs generated by \system with the API method name $\mathtt{readLine}$ as
a label. Names of variables of type $\mathtt{T}$ whose values are 
obtained from the environment are of the form $\mathtt{\$T}$.}\figlabel{example1}
\vspace{-0.1in}
\end{figure}

\subsection{Example}
Consider the label $\sX = (\sX_{\Calls}, \sX_{\Types},\Sx_{\Kws})$ where
$\sX_{\Calls} = \{\texttt{readLine}\}$ and $\sX_{\Types}$ and 
$\Sx_\Kws$ are empty.
\figref{example1}(a) shows a program that our best learner 
stochastically returns given this input. As we see,
this program indeed reads lines from a file, whose name is given by 
a special variable 
$\mathtt{\$String}$ that the code takes as input. It also handles
exceptions and closes the reader, 
even though these actions were not directly specified.

Although the program in \figref{example1}-(a) matches the label
well, failures do occur. Sometimes, the system generates a
program as in \figref{example1}-(b), which uses an {\tt
  InputStreamReader} rather than a {\tt FileReader}.  
%
It is possible to rule out this program by adding to the label. 
Suppose we amend $\sX_{Types}$ so that
$\sX_{Types} = \{ \texttt{FileReader} \}$. \system now tends to only generate programs that use {\tt
FileReader}. The variations then arise from different ways of handling
exceptions and constructing {\tt FileReader} objects (some programs
use a {\tt String} argument, while others use a {\tt File} object).
\figref{example2} in the 
appendix 
 shows two other top-five
programs returned on this input.


%% file: bayesian.tex

\newcommand{\Exp}{\mathsf{Exp}}

\newcommand{\sat}{\mathit{sat}}


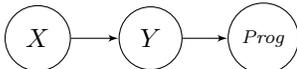
\begin{wrapfigure}{l}{1.5in}
\vspace{-0.5cm}
\begin{tikzpicture}

\tikzset{vertex/.style = {shape=circle,draw}}
\tikzset{edge/.style = {->,> = latex'}}

\node[vertex] (a) at  (-1.5,0) {$~X~$};
\node[vertex] (b) at  (0,0) {$~Y~$};
\node[vertex] (c) at  (1.5,0) {{\scriptsize $\Prog$}};
\draw[edge] (a) to (b);
\draw[edge] (b) to (c);
\end{tikzpicture}
\caption{Bayes net for $\Prog$, $X$, $Y$}\figlabel{bayesnet}
\vspace{-0.2cm}
\end{wrapfigure}

Our approach is to learn $g$ via maximum conditional likelihood
estimation (CLE).  That is, given a distribution family
$P (Prog | X, \theta)$ for a parameter set $\theta$, we choose
$\M^* = \argmax_{\M} \sum_i \log P(\Sprog_i \mid \Sx_i, \M)$.  Then,
$g(\sX) = \argmax_{\sProg} P(\sProg | \sX, \M^*)$.

The key innovation of our approach is that here, learning
happens at a higher level of abstraction than $(\sX_i, \sProg_i)$ pairs.  In practice, Java-like programs
contain many low-level details (for example, variable names and intermediate results) that can obscure
patterns in code.  Further, they contain complicated
semantic rules (for example, for type safety)
that are difficult to learn from data.  In contrast, these are relatively easy for a combinatorial,
syntax-guided program synthesizer \citep{sygus} to deal with.  However, synthesizers have a notoriously
difficult time figuring out the correct ``shape'' of a program (such as the placement of loops and conditionals), which
we hypothesize should be relatively easy for a statistical learner.

Specifically, our approach learns over 
{\em sketches}: tree-structured data that
capture key facets of program syntax. A sketch $\Sy$ does not
contain low-level variable names and operations, but carries information
about broadly shared facets of programs such as the types and API
calls. 
During generation,
a program synthesizer is used to generate programs from sketches produced by the learner.

Let the universe of all sketches be denoted
by $\mathbb{Y}$.
The sketch for a given program is computed by applying 
an {\em abstraction function}
$\alpha: \mathbb{P} \rightarrow \mathbb{Y}$. We call a sketch $\Sy$
{\em satisfiable}, and write $\sat(\Sy)$, if
$\alpha^{-1}(\Sy) \ne \emptyset$. The process of generating
(type-safe) programs given a satisfiable sketch $\Sy$ is
probabilistic, and captured by a {\em concretization distribution}
$P(\Prog \mid \Sy, \sat(\Sy))$. We 
require that for all programs $\Sprog$ and sketches $\Sy$ such that
$\sat(\Sy)$, we have $P(\Sprog \mid \Sy) \ne 0$ only if
$\Sy = \alpha(\Sprog).$

Importantly, the concretization distribution is fixed and chosen
heuristically. The alternative of learning this distribution from
source code poses difficulties: a single sketch can correspond to many
programs that only differ in superficial details, and deciding which
differences between programs are superficial and which are not
requires knowledge about program semantics. In contrast, our heuristic
approach utilizes known semantic properties of programming languages
like ours --- for example, that local variable names do not matter,
and that some algebraic expressions are semantically equivalent. This
knowledge allows us to limit the set of programs that we generate.

\begin{wrapfigure}{l}{2.5in} 
\vspace{-0.15in}
{\small 
$
\begin{array}{lll}
\Sy & ::= & \skipc ~|~ \callc~\id{Cexp} ~|~ \Sy_1; \Sy_2 ~|~\smallskip \\
& & \ifc~\id{Cseq}~\thenc~\Sy_1~\elsec~\Sy_2 ~|~\smallskip \\
& &  \whilec~\id{Cseq}~\doc~\Sy_1 ~|~ \tryc~\Sy_1~\id{Catch}
               \smallskip \\
\id{Cexp} & ::= & \tau_0.a(\tau_1,\dots, \tau_k) \smallskip \\
\id{Cseq} & ::= & \textrm{List of}~\id{Cexp} \smallskip \\
\id{Catch} & ::= & \catchc (\tau_1)~\Sy_1~\dots~\catchc(\tau_k)~\Sy_k
\end{array}
$
}
\vspace{-0.05in}
\caption{Grammar for sketches}\figlabel{sketchlang}

\end{wrapfigure}

Let us define a random variable $Y = \alpha(\Prog)$. 
We assume that the variables $X$,
$Y$ and $\Prog$ are related as in the Bayes net in \figref{bayesnet}. 
Specifically, given $Y$, $\Prog$ is
conditionally independent of $X$.
Further, let us assume a distribution family $P(Y | X, \theta)$ parameterized on $\theta$.

Let $\Sy_i =
\alpha(\Sprog_i)$, and note that $P(\Sprog_i | \Sy) \ne 0$ only if
$\Sy = \Sy_i$. Our problem now simplifies to 
{\em
  learning over sketches}, i.e., finding
\begin{eqnarray}\label{eqlearning}
\M^* & = & \argmax_\M \sum_i \log \sum_{\Sy: \sat(\Sy) }P(\Sprog_i|
           \Sy) P(\Sy | \Sx_i, \M) \nonumber\\
& = & \argmax_\M \sum_i \log P(\Sprog_i|
      \Sy_i) P(\Sy_i | \Sx_i, \M) = \argmax_\M \sum_i \log P(\Sy_i | \Sx_i, \M).
\end{eqnarray}

\subsection{Instantiation} 
\figref{sketchlang} shows the full grammar for sketches in our
implementation.
Here, $\tau_0, \tau_1,\dots$ range over a 
finite set of {\em API data types} that \lang programs can use. A data
type, akin to a Java class, is identified
with a finite set of {\em API method names} (including constructors),
and $a$ ranges over these names. 
Note that sketches do not contain 
constants or variable names. 

A full definition of the abstraction function for \lang appears in
Appendix~\ref{abs-appendix}. 
As an 
 example, API calls in \lang have the syntax
 ``$\callc~e.a(e_1,\dots, e_k)$'', where $a$ is an API method, the
expression $e$ evaluates to the object on which the method is called,
and the expressions $e_1,\dots, e_k$ evaluate to the arguments of the method call.
We abstract this call into an {\em abstract method
  call} ``$\callc~\tau.a(\tau_1,\dots, \tau_k)$'', where $\tau$ is the type of
$e$ and $\tau_i$ is the type of $e_i$. 
The keywords $\skipc$, $\whilec$, 
$\ifc$-$\thenc$-$\elsec$, and 
$\tryc$-$\catchc$ 
 preserve information about control flow and exception
 handling. Boolean conditions $\id{Cseq}$ are replaced by 
{\em abstract expressions}: {lists} whose elements abstract the API calls
in $\id{Cseq}$.

%% file: variational.tex
Now we describe our learning approach.
Equation~\ref{eqlearning} leaves us with the problem of computing $\argmax_\M \sum_i \log P(\Sy_i | \Sx_i, \M)$, when each $\sX_i$
is a label and $\sY_i$ is a sketch.
Our answer is to utilize an encoder-decoder and introduce a real vector-valued latent variable $Z$ 
to stochastically link labels and sketches: 
$P(\sY | \sX, \M) = \int_{\sZ \in \mathbb{R}^m} P(\sZ | \sX, \M) P (\sY | \sZ, \M) d\sZ.$

$P (Y | \sZ, \M)$ is realized as a probabilistic decoder mapping a vector-valued variable 
to a distribution over trees.  
We describe this decoder in Appendix~\ref{appendix-neural}.
As for $P(Z | \sX, \M)$, this distribution can, in principle, be
picked in any way we like. 
 In practice, because
both $P (Y | \sZ, \M)$ and
$P(Z | \sX, \M)$ have neural components with numerous parameters, we wish this distribution to regularize
the learner.
To provide this regularization, 
we assume a Normal $(\vec{0}, \textbf{I})$ prior on $Z$.  

Recall
that our labels are of the form
$\Sx = (\sX_{Calls}, \sX_{Types}, \Sx_{\Kws})$, where
$\sX_{\Calls}$, $\sX_{\Types}$, and $\Sx_{\Kws}$ are sets.  Assuming that the $j$-th elements
$\sX_{\Calls, j}$, $\sX_{\Types, j}$, and $\sX_{\Kws, j}$ of these sets are generated
independently, and assuming a function $f$ for encoding these
elements, let:
\begin{eqnarray*}
P(\sX | \sZ, \M) & = & \left( \prod_j \textrm{Normal} (f(\sX_{Calls,j}) | \sZ, \textbf{I} \sigma^2_{Calls}) \right) 
	\left( \prod_j \textrm{Normal} (f(\sX_{Types,j}) | \sZ,
          \textbf{I} \sigma^2_{Types}) \right) \\
& & \left( \prod_j \textrm{Normal} (f(\sX_{\Kws,j}) | \sZ,
          \textbf{I} \sigma^2_{\Kws}) \right).
\end{eqnarray*} 
That is, the encoded value of each $\sX_{\Types,j}$, $\sX_{\Calls,j}$ or $\sX_{\Kws,j}$
is sampled from a high-dimensional Normal distribution
centered at $\sZ$.  
If $f$ is 1-1 and onto with the set $\mathbb{R}^m$ then from
Normal-Normal conjugacy, we have:
$
P(\Sz | \Sx) = \textrm{Normal}\left( \Sz ~\big\lvert~  
  \frac{\bar{\mathsf{X}} }{1+n}, \frac{1}{1+n} \I \right) 
$, 
where
$$\bar{\mathsf{X}} =  
	\left( \sigma_{\Types}^{-2} \sum\limits_{j} f(\mathsf{X}_{\Types, j})  \right) + 
	\left( \sigma_{\Calls}^{-2} \sum\limits_{j}
          f(\mathsf{X}_{\Calls, j}) \right) + \left( \sigma_{\Kws}^{-2} \sum\limits_{j}
          f(\mathsf{X}_{\Kws, j}) \right) $$
and
$n = n_{\Types} \sigma_{\Types}^{-2} + n_{\Calls} \sigma_{\Calls}^{-2}
+ n_{\Kws} \sigma_{\Kws}^{-2}$.
Here, $n_{\Types}$ is the number of types supplied, and $n_{\Calls}$
and $n_{\Kws}$ are defined similarly.  

Note that this particular $P(Z | \Sx, \M)$ only follows directly
from the Normal $(\vec{0}, \textbf{I})$ prior on $Z$ and Normal likelihood 
$P(X | \sZ, \M)$ if the encoding function $f$ is 1-1 and onto.  However, even if $f$ is \emph{not} 1-1 and onto (as will be the case if
$f$ is implemented with a standard
feed-forward neural network)
we can still use this probabilistic encoder, and in practice we still tend to see the benefits of the regularizing prior on 
$\sZ$, with $P(Z)$ distributed approximately according to a unit Normal.
We call this type of encoder-decoder, with a single, Normally-distributed latent variable $Z$ linking the input and output,
a Gaussian encoder-decoder, or \modelname for short.

Now that we have chosen $P(X | \sZ, \M)$ and 
$P (Y | \sZ, \M)$, we must choose $\M$ to perform CLE. Note that:
\begin{align}
\sum_i \log P(\sY_i | \sX_i, \M) \nonumber 
&= \sum_i \log \int_{\sZ \in \mathbb{R}^m} P(\sZ | \sX_i, \M) P (\sY_i | \sZ, \M) d\sZ \nonumber = \sum_i \log \textbf{E}_{\sZ \sim P(Z | \sX_i, \M)} [P (\sY_i | \sZ, \M)] \nonumber \\
&\ge \sum_i \textbf{E}_{\sZ \sim P(Z | \sX_i, \M)} [\log P (\sY_i | \sZ, \M)] 
= \mathcal{L}(\M). \nonumber
\end{align}
\noindent where the $\ge$ holds due to Jensen's inequality.  Hence, $\mathcal{L}(\M)$ serves as a lower bound
on the log-likelihood, and so we can compute $\M^* = \argmax_{\M}
\mathcal{L}(\M)$ as a proxy for the CLE.  We maximize this lower bound
using stochastic gradient ascent; 
as $P(Z | \sX_i, \M)$ is Normal, we can use the 
re-parameterization 
trick common in variational auto-encoders \citep{kingma2013auto} while
doing so.
The parameter set $\theta$ contains all of the parameters of the
encoding function $f$ as well as $\sigma_{\Types}$, 
$\sigma_{\Calls}$, and $\sigma_{\Kws}$, and the parameters used in the decoding distribution funciton
$P (Y | \sZ, \M)$.

%% file: synthesis-short.tex
The final step in our algorithm is to ``concretize'' sketches into
 programs, following the distribution $P(\Prog | \Sy)$. Our method of
 doing so is a type-directed, stochastic search procedure that builds
 on combinatorial methods for program
 synthesis~\citep{Schkufza0A16,example-pldi15}. 

Given a sketch $\Sy$, our procedure performs a random walk in a space
of {\em partially concretized sketches} (PCSs). A PCS is a term
obtained by replacing some of the abstract method calls and expressions in a sketch by \lang method calls and \lang expressions. For example, the term ``$x_1.a(x_2); \tau_1.b(\tau_2)$'', which sequential composes an abstract method call to $b$ and a ``concrete'' method call to $a$, is a PCS.
The state of the procedure at the $i$-th point of the walk is 
a PCS $\Sh_i$. 
The initial state is $\Sy$.

Each state $\Sh$ has a set of {\em neighbors}
$\Next(\Sh)$. This set consists of all PCS-s
$\Sh'$ that are obtained by concretizing a single
abstract method call or expression in $\Sh$, using variable names in a way that
is consistent 
with the types of all API methods and declared variables
in $\Sh$.

The $(i + 1)$-th state in a walk is a sample from a predefined,
heuristically chosen distribution
$P(\Sh_{i+1} \mid \Sh_i,)$. The only
requirement on this distribution is that it assigns nonzero
probability to a state iff it belongs to
$\Next(\Sh_i)$. In practice, our implementation of this distribution
prioritizes programs that are simpler.
The random walk ends when it reaches a state $\Sh^*$ that
has no neighbors. If $\Sh^*$ is fully concrete (that is, an \lang
program), then the walk is successful and $\Sh^*$ is returned as a
sample. 
If not, the current walk is rejected, and a fresh walk is
started from the initial state. 
 
Recall that the concretization distribution $P(\Prog | \Sy)$ is only defined for sketches $\Sy$ that are satisfiable. Our concretization procedure does not {\em assume} that its input $\Sy$ is satisfiable. However, if $\Sy$ is not satisfiable, all random walks that it performs end with rejection, causing it to never terminate.

While the worst-case complexity of this procedure is exponential in
the generated programs,  
it performs well in practice because of our chosen
language of sketches. For instance, our search does not need
to discover the 
high-level structure of programs. 
Also, sketches specify the types of method arguments and return values, and 
this significantly limits the search space.

%% file: eval.tex
Now we present an empirical evaluation of the effectiveness of our
method. The experiments we describe utilize data from an online
repository of about 1500 Android apps~\citep{androiddrawer}. We decompiled the
APKs using JADX~\citep{JADX} to generate their source code. Analyzing about 100
million lines of code that were generated, we extracted 150,000 methods that
used Android APIs or the Java library. We then pre-processed all method
bodies to translate the code from Java to \lang, 
preserving names of relevant API
calls and data types as well as the high-level control flow. Hereafter, when we say ``program'' we refer to
an \lang program.

\begin{wrapfigure}{l}{2.1in}
{\footnotesize 
\vspace{-0.1in}
\begin{tabular}{|l|l|l|l|l|}
\hline
& {\scriptsize Min} & {\scriptsize Max} & {\scriptsize Median} &
                                                                 {\scriptsize Vocab} \\
\hline
$\Sx_\Calls$ &   1   &   9    &     2      & 2584  \\
\hline
$\Sx_\Types$ &   1   &   15    &     3       & 1521   \\
\hline
$\Sx_\Kws$ &    2  &    29  &       8      & 993  \\
\hline
$\Sx$ &    4  &    48  &       13      & 5098  \\
\hline
\end{tabular}
}
\caption{Statistics on labels}\figlabel{datastats}
\vspace{-0.1in}
\end{wrapfigure}

From each program, we extracted the sets $\Sx_{\Calls}$,
$\Sx_{\Types}$, and $\Sx_\Kws$ as well as a sketch
$\Sy$. Lacking separate natural language dscriptions for
programs, we defined keywords to be words obtained by splitting the
names of the API types and calls that the program uses, based on camel
case. For instance, the keywords obtained from the API call ${\tt readLine}$ are ``read'' and ``line''. As API method and types
in Java tend to be carefully named, these words often contain rich
information about what programs do.  \figref{datastats} gives some
statistics on the sizes of the labels in the data.  
From the extracted data, we randomly selected 10,000 programs to be in
the testing and validation data each.



\subsection{Implementation and training}
We implemented our approach in our tool called \system, using
TensorFlow~\citep{tensorflow} to implement the \modelname neural model, and the Eclipse
IDE for the abstraction from Java to the language of sketches and the
combinatorial concretization.
%
%

In all our experiments we performed cross-validation through grid search and
picked the best performing model.  Our
hyper-parameters for training the model are as follows. We used 64, 32 and 64
units in the encoder for API calls, types and keywords, respectively, and 128
units in the decoder. The latent space was 32-dimensional. We used a mini-batch
size of 50, a learning rate of 0.0006 for the Adam gradient-descent
optimizer~\citep{kingma2014adam}, and ran the training for 50 epochs.

The training was performed on an AWS ``p2.xlarge'' machine with an NVIDIA K80
GPU with 12GB GPU memory. As each sketch was broken down into a set of
production paths, the total number of data points fed to the model was around
700,000 per epoch. Training took 10 hours to complete.

\lstset{mathescape=true,numbers=none,basicstyle=\ttfamily\footnotesize}
\begin{wrapfigure}{l}{0.51\textwidth}
   \vspace{-0.2cm}
\includegraphics[scale=0.4]{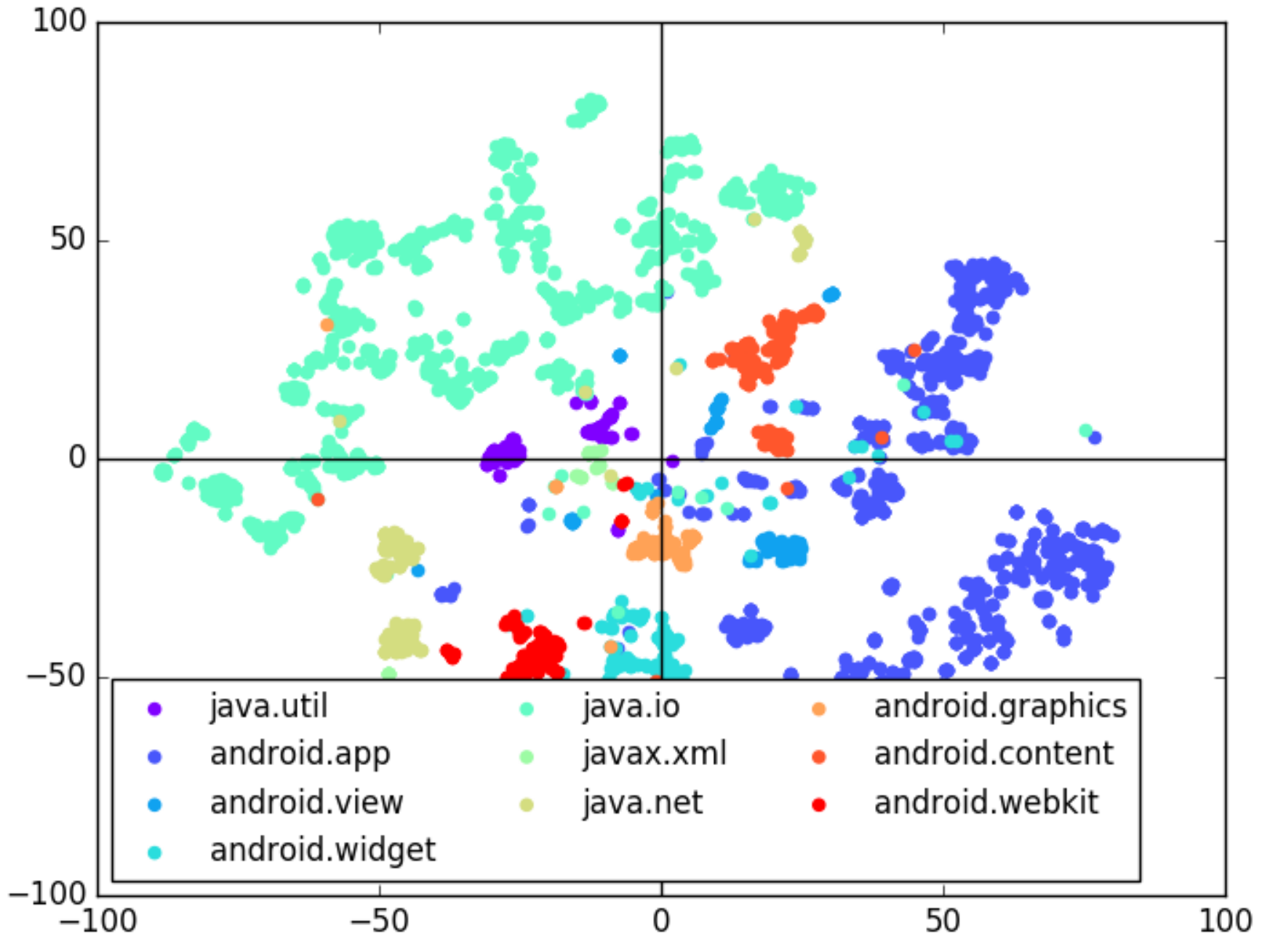}
  \vspace{-0.1in}
\caption{2-dimensional projection of latent space}
\figlabel{2dplot}
\end{wrapfigure}

\subsection{Clustering}
To visualize clustering in the 32-dimensional latent space, we provided labels
$\Sx$ from the testing data and sampled $\Sz$ from $P(Z|\Sx)$, and then used it to
sample a sketch from $P(Y|\Sz)$. We then used t-SNE~\citep{maaten2008tsne} to
reduce the dimensionality of $\Sz$ to 2-dimensions, and labeled each point with
the API used in the sketch $\Sy$.
\figref{2dplot} shows this 2-dimensional space, where each label has been coded
with a different color. It is immediately apparent from the plot that the model
has learned to cluster the latent space neatly according to different APIs.
Some APIs such as ${\tt java.io}$ have several modes, and we noticed separately
that each mode corresponds to different usage scenarios of the API, such as
reading versus writing in this case.

\subsection{Accuracy}
To evaluate prediction accuracy, we provided labels from the testing data to our
model, sampled sketches from the distribution $P(Y|\Sx)$ and concretized each
sketch into an \lang program using our combinatorial search. We then measured the
number of test programs for which a program that is equivalent to the expected
one appeared in the top-10 results from the model.

As there is no universal metric to measure program equivalence (in fact, it is
an undecidable problem in general), we used several metrics to approximate the
notion of equivalence. We defined the following metrics on the top-10 programs
predicted by the model:

\begin{itemize}
  \item[M1.] This binary metric measures whether the expected program
      appeared in a syntactically equivalent form in the results.  Of course, an
      impediment to measuring this is that the names of variables used in the
      expected and predicted programs may not match. It is neither reasonable
      nor useful for any model of code to learn the exact variable names in the training
      data. Therefore, in performing this equivalence check, we abstract away
      the variable names and compare the rest of the program's Abstract Syntax
      Tree (AST) instead.

  \item[M2.] This metric measures the minimum {\em Jaccard
      distance} between the sets of sequences of API calls made
      by the expected and predicted programs.  It is a measure of how close to
      the original program were we able to get in terms of sequences of API
      calls.

  \item[M3.] Similar to metric M2, this metric measures the minimum Jaccard
      distance between the sets of API calls in the expected and predicted
      programs.  

  \item[M4.] This metric computes the minimum absolute difference between the
      number of statements in the expected and sampled programs, as a ratio of
      that in the former.

  \item[M5.] Similar to metric M4, this metric computes the minumum absolute
      difference between the number of control structures in the expected and
      sampled programs, as a ratio of that in the former. Examples of control
      structures are branches, loops, and try-catch statements.
\end{itemize}

\subsection{Partial Observability}
To evaluate our model's ability to predict programs given a small amount of
information about its code, 
we varied the 
fraction of the set of API calls, types, and keywords 
provided as input from the testing data.
We experimented with 75\%, 50\% and 25\% observability in the testing
data; the median number of items in a label in these
cases were 
9, 6, and 2, respectively.

\subsection{Competing Models}
In order to compare our model with state-of-the-art conditional generative
models, we implemented the Gaussian Stochastic Neural Network (\GSNN) presented
by~\citep{GSNN}, using the same tree-structured decoder as the \modelname.
There are two main differences: (i) the \GSNN's decoder is also conditioned
directly on the input label $\Sx$ in addition to $\Sz$, which we accomplish by
concatenating its initial state with the encoding of $\Sx$, (ii) the \GSNN loss
function has an additional KL-divergence term weighted by a hyper-parameter
$\beta$. We subjected the \GSNN to the same training and cross-validation
process as our model. In the end, we selected a model that happened to have very
similar hyper-parameters as ours, with $\beta=0.001$.

\subsection{Evaluating Sketches}
In order to evaluate the effect of sketch learning for program generation, we
implemented and compared with a model that learns directly over programs.
Specifically, the neural network structure is exactly the same as ours, except
that instead of being trained on production paths in the sketches, the model is
trained on production paths in the ASTs of the \lang programs. We selected
a model that had more units in the decoder (256) compared to
our model (128), as the \lang grammar is more complex than the grammar of
sketches. We also implemented a similar \GSNN model to train over \lang ASTs
directly.

\begin{figure*}
  \begin{tabular}{ll}
    \begin{minipage}{0.45\textwidth}
{\scriptsize
    \begin{tabular}{|l|r|r|r|r|}
        \hline
        {\bf Model} & \multicolumn{4}{|c|}{{\bf Input Label Observability}} \\ \cline{2-5}
        & $100\%$ & $75\%$ & $50\%$ & $25\%$ \\ \hline
        \modelname-\lang  & $0.13$ & $0.09$ & $0.07$ & $0.02$ \\
        \GSNN-\lang & $0.07$ & $0.04$ & $0.03$ & $0.01$ \\
        \modelname-Sk   & ${\bf 0.59}$ & ${\bf 0.51}$ & ${\bf 0.44}$ & ${\bf 0.21}$ \\
        \GSNN-Sk  & $0.57$ & $0.48$ & $0.41$ & $0.18$ \\ \hline
    \end{tabular} \\ \\
}
{\footnotesize    (a) M1. Proportion of test programs for which the expected AST appeared in
    the top-10 results.}
    \end{minipage}
    &
    \begin{minipage}{0.45\textwidth}
{\scriptsize
\begin{tabular}{|l|r|r|r|r|}
        \hline
        {\bf Model} & \multicolumn{4}{|c|}{{\bf Input Label Observability}} \\ \cline{2-5}
        & $100\%$ & $75\%$ & $50\%$ & $25\%$ \\ \hline
        \modelname-\lang  & $0.82$ & $0.87$ & $0.89$ & $0.97$ \\
        \GSNN-\lang & $0.88$ & $0.92$ & $0.93$ & $0.98$ \\
        \modelname-Sk   & ${\bf 0.34}$ & ${\bf 0.43}$ & ${\bf 0.50}$ & ${\bf 0.76}$ \\
        \GSNN-Sk  & $0.36$ & $0.46$ & $0.53$ & $0.78$ \\ \hline
    \end{tabular} \\ \\
}
{\footnotesize    (b) M2. Average minimum Jaccard distance on the set of sequences of API methods called in
    the test program vs the top-10 results.}
    \end{minipage}
    \\ \\
    \begin{minipage}{0.45\textwidth}
{\scriptsize 
   \begin{tabular}{|l|r|r|r|r|}
        \hline
        {\bf Model} & \multicolumn{4}{|c|}{{\bf Input Label Observability}} \\ \cline{2-5}
        & $100\%$ & $75\%$ & $50\%$ & $25\%$ \\ \hline
        \modelname-\lang  & $0.52$ & $0.58$ & $0.61$ & $0.77$ \\
        \GSNN-\lang & $0.59$ & $0.64$ & $0.68$ & $0.83$ \\
        \modelname-Sk   & ${\bf 0.11}$ & ${\bf 0.17}$ & ${\bf 0.22}$ & ${\bf 0.50}$ \\
        \GSNN-Sk  & $0.13$ & $0.19$ & $0.25$ & $0.52$ \\ \hline
    \end{tabular} \\ \\
}
{\footnotesize    (c) M3. Average minimum Jaccard distance on the set of API methods called in
    the test program vs the top-10 results.}
    \end{minipage}
    & 
    \begin{minipage}{0.45\textwidth}
{\scriptsize 
    \begin{tabular}{|l|r|r|r|r|}
        \hline
        {\bf Model} & \multicolumn{4}{|c|}{{\bf Input Label Observability}} \\ \cline{2-5}
        & $100\%$ & $75\%$ & $50\%$ & $25\%$ \\ \hline
        \modelname-\lang  & $0.49$ & $0.47$ & $0.46$ & $0.46$ \\
        \GSNN-\lang & $0.52$ & $0.49$ & $0.49$ & $0.53$ \\
        \modelname-Sk   & ${\bf 0.05}$ & ${\bf 0.06}$ & ${\bf 0.06}$ & ${\bf 0.09}$ \\
        \GSNN-Sk  & ${\bf 0.05}$ & ${\bf 0.06}$ & ${\bf 0.06}$ & ${\bf 0.09}$ \\ \hline
    \end{tabular} \\ \\
}
{\footnotesize    (d) M4. Average minimum difference between the number of statements in the
    test program vs the top-10 results.}
    \end{minipage}
    \\ \\
    \begin{minipage}{0.45\textwidth}
{\scriptsize  
   \begin{tabular}{|l|r|r|r|r|}
        \hline
        {\bf Model} & \multicolumn{4}{|c|}{{\bf Input Label Observability}} \\ \cline{2-5}
        & $100\%$ & $75\%$ & $50\%$ & $25\%$ \\ \hline
        \modelname-\lang  & $0.31$ & $0.30$ & $0.30$ & $0.34$ \\
        \GSNN-\lang & $0.32$ & $0.31$ & $0.32$ & $0.39$ \\
        \modelname-Sk   & ${\bf 0.03}$ & ${\bf 0.03}$ & ${\bf 0.03}$ & $0.04$ \\
        \GSNN-Sk  & ${\bf 0.03}$ & ${\bf 0.03}$ & ${\bf 0.03}$ & ${\bf 0.03}$ \\ \hline
    \end{tabular} \\ \\
}
 {\footnotesize   (e) M5. Average minimum difference between the number of control structures
    in the test program vs the top-10 results.}
    \end{minipage}
    &
    \begin{minipage}{0.45\textwidth}
{\scriptsize  
   \begin{tabular}{|l|r|r|r|r|r|}
            \hline
            {\bf Model} & \multicolumn{5}{|c|}{{\bf Metric}} \\ \cline{2-6}
                        & M1 & M2 & M3 & M4 & M5 \\ \hline
            \modelname-\lang  & $0.02$ & $0.97$ & $0.71$ & $0.50$ & $0.37$ \\
            \GSNN-\lang & $0.01$ & $0.98$ & $0.74$ & $0.51$ & $0.37$ \\
            \modelname-Sk   & ${\bf 0.23}$ & ${\bf 0.70}$ & ${\bf 0.30}$ & ${\bf 0.08}$ & ${\bf 0.04}$ \\
            \GSNN-Sk  & $0.20$ & $0.74$ & $0.33$ & ${\bf 0.08}$ & ${\bf 0.04}$ \\ \hline
        \end{tabular} \\ \\
}
{\footnotesize    (f) Metrics for $50\%$ obsevability evaluated only on unseen data}
    \end{minipage}
  \end{tabular}
  \caption{Accuracy of different models on testing data. \modelname-\lang and \GSNN-\lang
  are baseline models trained over \lang ASTs, \modelname-Sk and \GSNN-Sk are models
trained over sketches.}
  \figlabel{metrics}
\vspace{-0.2in}
\end{figure*}

\vspace{2mm}
\figref{metrics} shows the collated results of this evaluation, where each entry
computes the average of the corresponding metric over the 10000 test programs.
It takes our model about 8 seconds, on average, to generate and rank 10
programs.

When testing models that were trained on \lang ASTs, namely the \modelname-\lang and \GSNN-\lang models,
we observed that out of a total of 87,486 \lang ASTs sampled from the two models,
2525 (or 3\%) ASTs were not even well-formed, i.e., they would not pass a
parser, and hence had to be discarded from the metrics. This number is 0 for the
\modelname-Sk and \GSNN-Sk models, meaning that all \lang ASTs that were obtained by
concretizing sketches were well-formed.

In general, one can observe that the \modelname-Sk model performs best overall, with
\GSNN-Sk a reasonable alternative.  We hypothesize that the reason \modelname-Sk performs
slightly better is the regularizing prior on $Z$; since the GSNN has a direct link from
$X$ to $Y$, it can choose to ignore this regularization. We would classify both these models as suitable
for conditional program generation. However, the other two models \modelname-\lang and
\GSNN-\lang perform quite worse, showing that sketch learning is key in addressing
the problem of conditional program generation.

\subsection{Generalization}
To evaluate how well our model generalizes to unseen data, we gather a subset of
the testing data whose data points, consisting of label-sketch pairs
$(\Sx,\Sy)$, never occurred in the training data. We then evaluate the same
metrics in \figref{metrics}(a)-(e), but due to space reasons we focus on the 50\%
observability column. \figref{metrics}(f) shows the results of this evaluation
on the subset of 5126 (out of 10000) unseen test data points. The metrics
exhibit a similar trend, showing that the models based on sketch learning are
able to generalize much better than the baseline models, and that the \modelname-Sk
model performs the best.

%% file: relwork.tex

{\em Unconditional}, corpus-driven
generation of programs has been studied before
\citep{ICML:MT14,idiomsFSE14,bielik2016phog}, as has the generation of
code snippets conditioned on a {\em context} into which the snippet is
merged \citep{nguyenFSE13,raychev2014code,nguyenICSE15}. These prior efforts
often use models like $n$-grams~\citep{nguyenFSE13} and
recurrent neural networks~\citep{raychev2014code} that are primarily suited
to the generation of straight-line programs; almost universally, they
cannot guarantee semantic properties of generated programs. Among prominent
exceptions, 
\citet{ICML:MT14} use
log-bilinear tree-traversal models, a class of probabilistic pushdown
automata, for program generation. \citet{bielik2016phog} study a
generalization of probabilistic grammars known as probabilistic
higher-order grammars.  Like our work, these papers address the
generation of programs that satisfy rich constraints such as the
type-safe use of names. In principle, one could replace our decoder
and the combinatorial concretizer, which together form an
unconditional program generator, with one of these models. However,
given our experiments, doing so is unlikely to lead to good
performance in the end-to-end problem of conditional program
generation.

%
%

There is a line of existing work considering the generation of programs
from text \citep{yin2017syntactic,ling2016latent,rabinovich2017abstract}.  These papers use decoders similar to the one
used in \system, and since they are solving the text-to-code problem, they utilize attention mechanisms not found in \system.
Those attention mechanisms could be particularly
useful were \system extended to handle natural language evidence.
The fundamental difference between these works and \system, however, is the level of abstraction at which learning takes place.
These papers attempt to translate text directly into code, whereas \system uses neural methods to produce higher-level sketches 
that are translated into program code using symbolic methods.  This two-step code generation process is central to \system.  It 
ensures key semantic properties of the generated code (such as type safety) and by abstracting away from the learner many lower-level details, it
may make learning easier. 
We have given experimental evidence that this 
approach can give better results than translating directly into code.

\citet{kusner2017grammar} propose a variational autoencoder for context-free
grammars. As an auto-encoder, this model is generative, but it is not
a conditional model such as ours. In their application of 
synthesizing molecular structures, given a particular molecular
structure, their model can be used to search the latent space for
similar valid structures. In our setting, however, we are not given a
sketch but only a label for the sketch, and our task is learn a
conditional model that can predict a whole sketch given a label.

Conditional program generation is closely related to {\em program
  synthesis}~\citep{gulwani-survey}, the problem of producing programs
  that satisfy a given semantic specification. The programming
  language community has studied this problem thoroughly using
  the tools of combinatorial search and symbolic reasoning~\citep{sygus,armandosketching,example-popl11,example-pldi15}.
  A common tactic in this literature is to put syntactic limitations on
  the space of feasible programs~\citep{sygus}. This is done either by
  adding a human-provided sketch to a problem
  instance~\citep{armandosketching}, or by restricting synthesis to a
  narrow DSL~\citep{example-popl11,flashmeta}.

  A recent body of work has developed neural approaches to program
  synthesis.  Terpret~\citep{gaunt2016terpret} and Neural
  Forth~\citep{riedel2016programming} use neural learning over a set
  of user-provided examples to complete a user-provided sketch.  In
  neuro-symbolic synthesis~\citep{parisotto2016neuro} and
  RobustFill~\citep{robustfill}, a neural architecture is used to
  encode a set of input-output examples and decode the resulting
  representation into a Flashfill program.
  DeepCoder~\citep{balog2016deepcoder} uses neural techniques to speed
  up the synthesis of Flashfill~\citep{inductive-realworld} programs.

These efforts differ from
  ours in goals as well as methods.  Our problem is simpler, as it is
  conditioned on syntactic, rather than semantic, facets of
  programs. This allows us to generate programs in a complex 
  programming language over a large number of data types
  and API methods, without needing a human-provided sketch. 
The key methodological difference between our work and symbolic
program synthesis lies in our use of data, which allows
us to generalize from a very small amount of specification. 
Unlike our approach, most neural approaches to program synthesis do
not 
combine learning and combinatorial
  techniques.  The prominent exception is Deepcoder~\citep{balog2016deepcoder},
  whose relationship with our work was discussed in \secref{intro}.

%% file: conc.tex
We have given a method for generating type-safe programs in a
Java-like language, given a label containing a small amount of
information about a program's code or metadata. Our main idea is to
learn a model that can predict {\em sketches} of programs relevant to
a label. The predicted sketches are concretized into code using
combinatorial techniques. We have implemented our ideas in \system, a
system for the generation of API-heavy code. Our experiments
indicate that the system can often generate complex method bodies 
from just a few tokens, and that learning at the level of sketches is
key to performing such generation effectively.

An important distinction between our work and classical program
synthesis is that our generator is conditioned on uncertain, syntactic
information about the target program, as opposed to hard constraints
on the program's semantics. Of course, the programs that we generate
are type-safe, and therefore guaranteed to satisfy certain semantic
constraints. However, these constraints are invariant across
generation tasks; in contrast, traditional program synthesis permits
instance-specific semantic constraints. Future work will seek to
condition program generation on syntactic labels {\em as well as}
semantic constraints. As mentioned earlier, learning correlations between the syntax
and semantics of programs written in complex languages is difficult. However,
the approach of first generating and then concretizing a sketch could
reduce this difficulty: sketches could be generated using a {\em
limited amount of} semantic information, and the concretizer could use
logic-based techniques~\citep{sygus,gulwani-survey} to ensure that the
programs synthesized from these sketches match the semantic
constraints exactly. A key challenge here would be to calibrate the
amount of semantic information on which sketch generation is
conditioned.


%% file: overview-appendix.tex
\lstset{mathescape=false,numbers=none,basicstyle=\ttfamily\scriptsize}
\begin{figure}[t]
\begin{tabular}{cc}
\begin{minipage}{0.52\textwidth}
\begin{lstlisting}[language=Java, basicstyle=\ttfamily\scriptsize]
 String s;
 BufferedReader br;
 FileReader fr;
 try {
  fr = new FileReader($String);
  br = new BufferedReader(fr);
  while ((s = br.readLine()) != null) {}
  br.close();
 } catch (FileNotFoundException _e) {
   _e.printStackTrace();
 } catch (IOException _e) {
   _e.printStackTrace(); 
 }
\end{lstlisting}
\end{minipage}
&
\begin{minipage}{0.47\textwidth}
\begin{lstlisting}[language=Java, basicstyle=\ttfamily\scriptsize]
 String s;
 BufferedReader br;
 FileReader fr; 
 try {
  fr = new FileReader($File);
  br = new BufferedReader(fr);
  while ((s = br.readLine()) != null){}
  br.close();
 } catch (FileNotFoundException _e){
 } catch (IOException _e){
 }
\end{lstlisting}
\end{minipage}
\\ (a) & (b) \\
\end{tabular}
\hfill
\vspace{-0.1in}
\caption{Programs generated in a typical run of \system, given the API method name {\tt readLine}
    and the type {\tt FileReader}.}\figlabel{example2}
\end{figure}


%% file: aml-appendix.tex
\begin{wrapfigure}{l}{2.9in}
\vspace{-0.1in}
{\small 
$
\begin{array}{lll}
\Sprog & ::= & \skipc ~|~ \Sprog_1; \Sprog_2 ~|~ \callc~\id{Call} ~|~
  \\
& & \declc~x = \id{Call} ~|~
               \smallskip \\
& & \ifc~\id{Exp}~\thenc~\Sprog_1~\elsec~\Sprog_2 ~|~
               \smallskip \\
& & \whilec~\id{Exp}~\doc~\Sprog_1 ~|~ \tryc~\Sprog_1~\id{Catch}
                     \smallskip \\
\id{Exp} & ::= & \id{Sexp} ~|~\id{Call} ~|~\letc~x = \id{Call}: \id{Exp}_1
                 \smallskip \\ 
\id{Sexp} & ::= & c ~|~ x \smallskip \\
\id{Call} & ::= & \id{Sexp_0}.a(\id{Sexp}_1,\dots, \id{Sexp}_k)
                  \smallskip \\ 
\id{Catch} & ::= & \catchc (x_1)~\Sprog_1~\dots~\catchc(x_k)~\Sprog_k
\end{array}
$ 
}
\caption{Grammar for \lang}\figlabel{grammar}
\vspace{-0.1in}
\end{wrapfigure}

\lang is a core language that is designed to
capture the essence of API usage in Java-like languages. Now we
present this language. 

\lang uses a finite set of {\em API data types}. A type is identified
with a finite set of {\em API method names} (including constructors);
the type for which this set is empty is said to be {\em void}.  Each
method name $a$ is associated with a {\em type signature}
$(\tau_1,\dots,\tau_k) \rightarrow \tau_0$, where
$\tau_1,\dots,\tau_k$ are the method's input types and $\tau_0$ is its
return type. A method for which $\tau_0$ is void is interpreted to not
return a value. Finally, we assume predefined universes of
{constants} and {variable names}.

The grammar for \lang is as in \figref{grammar}.  Here, $x, x_1,\dots$
are variable names, $c$ is a constant, and $a$ is a method name. The
syntax for programs $\Sprog$ includes method calls, loops, branches,
statement sequencing, and exception handling. We use variables to feed
the output of one method into another, and the keyword $\declc$ to
store the return value of a call in a fresh variable. $\id{Exp}$
stands for (object-valued) expressions, which include constants,
variables, method calls, and let-expressions such as
``$\letc~x = \id{Call}:~\id{Exp}$", which stores the return value of a
call in a fresh variable $x$, then uses this binding to evaluate the
expression $\id{Exp}$. (Arithmetic and relational operators are
assumed to be encompassed by API methods.) 


The operational semantics and type system for \lang are standard, and
consequently, we do not describe these in detail. 

%% file: abs-appendix.tex
\begin{figure}
{\footnotesize
\begin{eqnarray*}
\alpha(\skipc) & = & \skipc \\
\alpha(\callc~\id{Sexp}_0.a(\id{Sexp}_1,\dots, \id{Sexp}_k)) & = & \callc~\tau_0.a(\tau_1,\dots,\tau_k) ~~ \textrm{where $\tau_i$ is the type of $\id{Sexp}_i$}\\
\alpha(\Sprog_1; \Sprog_2) & = & \alpha(\Sprog_1); \alpha(\Sprog_2) \\
\alpha(\letc~x = \id{Sexp}_0.a(\id{Sexp}_1,\dots, \id{Sexp}_k)) & = &  \callc~\tau_0.a(\tau_1,\dots,\tau_k) ~~ \textrm{where $\tau_i$ is the type of $\id{Sexp}_i$}\\
\alpha(\ifc~\Exp~\thenc~\Sprog_1~\elsec~\Sprog_2) & = & 
  \ifc~\alpha(\Exp)~\thenc~\alpha(\Sprog_1)~\elsec~\alpha(\Sprog_2)
  \\
\alpha(\whilec~\Exp~\doc~\Sprog) & = & \whilec~\alpha(\id{{Cond}})~\doc~\alpha(\Sprog)\\
\alpha(\tryc~\Sprog~\catchc(x_1)~\Sprog_1~\dots~\catchc(x_k)~\Sprog_k) & = &
                                             \tryc~\alpha(\Sprog)~\\
  & &~~\catchc(\tau_1)~\alpha(\Sprog_1) \dots~\catchc(\tau_k)~\alpha(\Sprog_k) 
                                                                                \\
& & \textrm{where $\tau_i$ is the type of $x_i$} \medskip \\
\alpha(\Exp) & = & [~] \textrm{ if $\Exp$ is a
                        constant or variable name} \\
\alpha(\id{Sexp}_0.a(\id{Sexp}_1,\dots, \id{Sexp}_k)) & = & [\tau_0.a(\tau_1,\dots,\tau_k)]
                                    ~~~\textrm{where $\tau_i$ is the type of $\id{Sexp}_i$} \\
\alpha(\letc~x = \id{Call}: \Exp_1) & = &
                                                 \mathit{append}(\alpha(\id{Call}), \alpha(\Exp_1)) 
\end{eqnarray*}
}
\vspace{-0.2in}
\caption{The abstraction function $\alpha$.}\figlabel{abstraction}
\vspace{-0.1in}
\end{figure}

We define
the abstraction function $\alpha$ for the \lang language in \figref{abstraction}.

%% file: encoder.tex
In this section we present the details of the neural networks used by \system.

\subsection{The Encoder}
\label{subsec:enc}
The task of the neural encoder is to implement the encoding function $f$ for
labels, which accepts an element from a label, say $\Sx_{\Calls, i}$ as input
and maps it into a vector in $d$-dimensional space, where $d$ is the
dimensionality of the latent space of $Z$.
%
To achieve this, we first convert each element $\Sx_{\Calls,i}$ into its one-hot vector
representation, denoted $\Sx'_{\Calls,i}$.  Then, let
$h$ be the number of neural hidden units in the encoder for API calls, and let
${\bf W}_h \in \mathbb{R}^{|\Calls| \times h}$, ${\bf b}_h \in \mathbb{R}^{h}$, ${\bf W}_d \in
\mathbb{R}^{h \times d}$, ${\bf b}_d \in \mathbb{R}^{d}$ be real-valued weight and bias matrices
of the neural network. The encoding function $f(\Sx_{\Calls,i})$ can be defined as follows:
$$f(\Sx_{\Calls,i}) = {\sf tanh}(({\bf W}_h ~.~ \Sx'_{\Calls,i} + b_h) ~.~ {\bf W}_d + b_d)$$
where ${\sf tanh}$ is a non-linearity defined as ${\sf tanh}(x) = \frac{1 - e^{-2x}}{1 + e^{-2x}}$.
This would map any given API call into a $d$-dimensional real-valued vector. The values
of entries in the matrices ${\bf W}_h$, $b_h$, ${\bf W}_d$ and $b_d$ will be learned during
training. The encoder for types can be defined analogously, with its own set of
matrices and hidden state.

%% file: neural.tex
\subsection{The Decoder}
\newcommand{\codefont}[1]{\mbox{\small{\texttt{#1}}}}
\newcommand{\rootnode}{\mbox{{\sf root}}}
\newcommand{\trynode}{\mbox{{\sf try}}}
\newcommand{\catchnode}{\mbox{{\sf catch}}}
\newcommand{\loopnode}{\mbox{{\sf while}}}
\newcommand{\conditionnode}{\mbox{{\sf condition}}}
\newcommand{\bodynode}{\mbox{{\sf do}}}
\newcommand{\skipnode}{\mbox{{\sf skip}}}

\begin{figure}
\begin{center}   
\includegraphics[scale=0.4]{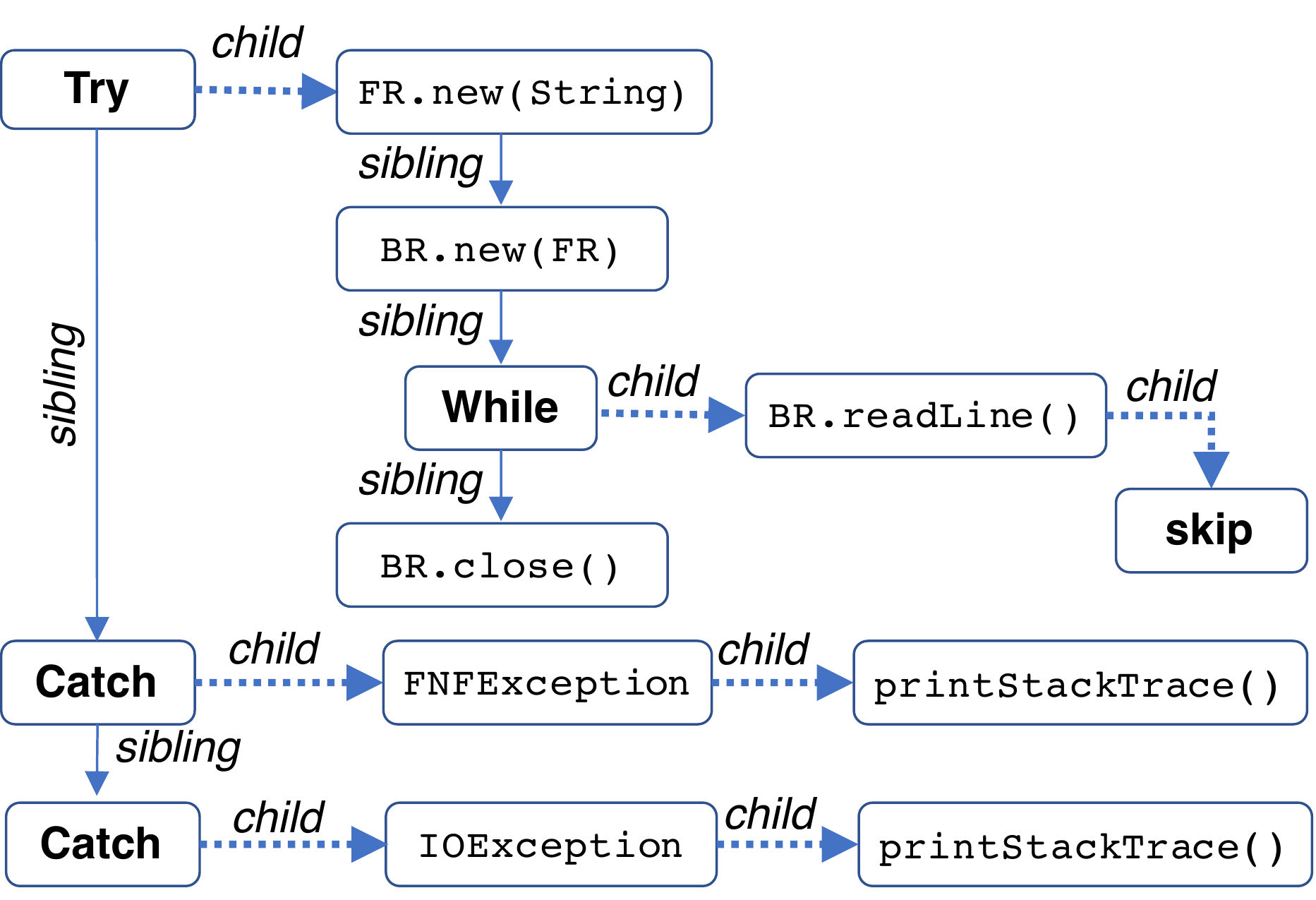}
\end{center}
\caption{Tree representation of the sketch in \figref{example2}(a)}
    \figlabel{decoder-tree}
\vspace{-0.2in}
\end{figure}

The task of the neural decoder is to implement the sampler for $\Sy \sim P(Y|\Sz)$.  This is
implemented recursively via repeated samples of production rules $\Sy_i$ in the grammar of sketches,
drawn as $\Sy_i \sim P(Y_i|\textbf{Y}_{i-1},\Sz)$, where $\textbf{Y}_{i-1} =
\Sy_1,\ldots,\Sy_{i-1}$.  The generation of each $\Sy_i$ requires the generation of a new ``path''
from a series of previous ``paths'', where each path corresponds to a series of production rules
fired in the grammar.

As a sketch is tree-structured, we use a top-down tree-structured recurrent neural network
similar to~\citet{treeRNN}, which we elaborate in this section. First, similar to the notion of a
``dependency path'' in~\citet{treeRNN}, we define a {\em production path} as a sequence of pairs
$\langle (v_1, e_1), (v_2, e_2), \ldots, (v_k, e_k) \rangle$ where $v_i$ is a node in the sketch
(i.e., a term in the grammar) and $e_i$ is the type of edge that connects $v_i$ with $v_{i+1}$. Our
representation has two types of edges: $sibling$ and $child$. A sibling edge connects two nodes at
the same level of the tree and under the same parent node (i.e., two terms in the RHS of the
same rule). A child edge connects a node with another that is one level deeper in the tree
(i.e., the LHS with a term in the RHS of a rule). We consider a
sequence of API calls connected by sequential composition as siblings.  The root of the entire tree
is a special node named \rootnode, and so the first pair in all production paths is $(\rootnode,
child)$. The last edge in a production path is irrelevant ($\cdot$) as it does not connect the node
to any subsequent nodes.

As an example, consider the sketch in \figref{example2}(a), whose representation as a tree for the
decoder is shown in \figref{decoder-tree}. For brevity, we use $s$ and $c$ for
$sibling$ and $child$ edges respectively, abbreviate some classnames with uppercase letters in
their name, and omit the first pair $(\rootnode, c)$ that occurs in all paths. There are
four production paths in the tree of this sketch:

\begin{enumerate}
  \item $(\tryc,c), (\codefont{FR.new(String)},s), (\codefont{BR.new(FR)},s),
      (\whilec,c), (\codefont{BR.readLine()},c), (\skipc,\cdot)$
  \item $(\tryc,c), (\codefont{FR.new(String)},s), (\codefont{BR.new(FR)},s),
      (\whilec,s), (\codefont{BR.close()},\cdot)$
  \item $(\tryc,s), (\catchc,c), (\codefont{FNFException},c),
      (\codefont{T.printStackTrace()},\cdot)$
  \item $(\tryc,s), (\catchc,s), (\catchc,c), (\codefont{IOException},c),
      (\codefont{T.printStackTrace()},\cdot)$
\end{enumerate}

Now, given a $\Sz$ and a sequence of pairs ${\bf Y}_i = \langle (v_1, e_1), \ldots, (v_i,
e_i) \rangle$ along a production path, the next node in the path is assumed to be dependent solely
on $\Sz$ and ${\bf Y}_i$. Therefore, a single inference step of the decoder computes the
probability $P(v_{i+1} | {\bf Y}_i, \Sz)$.  To do this, the decoder uses two RNNs, one for each
type of edge $c$ and $s$, that act on the production pairs in ${\bf Y}_i$. First,
all nodes $v_i$ are converted into their one-hot vector encoding, denoted $v_i'$.

Let $h$ be the number of hidden units in the decoder, and $|G|$ be the size of the decoder's output
vocabulary, i.e., the total number of terminals and non-terminals in the grammar of sketches. Let
${\bf W}_h^e \in \mathbb{R}^{h \times h}$ and ${\bf b}_h^e \in \mathbb{R}^{d}$ be the decoder's
hidden state weight and bias matrices, ${\bf W}_v^e \in \mathbb{R}^{|G| \times h}$ and ${\bf
b}_v^e \in \mathbb{R}^{h}$ be the input weight and bias matrices, and ${\bf W}_y^e \in \mathbb{R}^{h
\times |G|}$ and ${\bf b}_y^e \in \mathbb{R}^{|G|}$ be the output weight and bias matrices, where
$e$ is the type of edge: either $c$ (child) or $s$ (sibling).
We also use ``lifting'' matrices ${\bf W}_l \in \mathbb{R}^{d \times h}$ and ${\bf b}_l \in
\mathbb{R}^h$, to lift the $d$-dimensional vector $\Sz$ onto the (typically) higher-dimensional
hidden state space $h$ of the decoder.

Let $h_i$ and $y_i$ be the hidden state and output of the network at time point $i$. We compute
these quantities as given in \figref{decoder-eqns}, where
\begin{figure*}
\begin{align}
    h_0 &= {\bf W}_l . \Sz + {\bf b}_l
    &
    y_i &= \big \{
        \begin{tabular}{ll}
        ${\sf softmax}({\bf W}_y^c ~.~ h_i + {\bf b}_y^c)$ & if $e_i = child$ \\
        ${\sf softmax}({\bf W}_y^s ~.~ h_i + {\bf b}_y^s)$ & if $e_i = sibling$ \\
        \end{tabular}
    \label{eqn:rnnoutput} \\
    h_i^c &= {\bf W}_h^c ~.~ h_{i-1} + {\bf b}_h^c + {\bf W}_v^c ~.~ v_i' + {\bf b}_v^c &
    & \text{where} \nonumber \\
    h_i^s &= {\bf W}_h^s ~.~ h_{i-1} + {\bf b}_h^s + {\bf W}_v^s ~.~ v_i' + {\bf b}_v^s &
    &{\sf tanh}(x) = \frac{1 - e^{-2x}}{1 + e^{-2x}} \text{\hspace{0.2cm}and} \nonumber \\
    h_i &= \big \{
        \begin{tabular}{ll}
        ${\sf tanh}(h_i^c)$ & if $e_i = child$ \nonumber \\
        ${\sf tanh}(h_i^s)$ & if $e_i = sibling$ \nonumber 
        \end{tabular} &
    &{\sf softmax}({\bf x})_j = \frac{e^{x_j}}{\sum_{k=1}^K e^{x_k}} \text{\hspace{0.2cm}for } j \in 1 \ldots K \nonumber 
            \\ \nonumber
\end{align}
\caption{Computing the hidden state and output of the decoder}
\figlabel{decoder-eqns}
\end{figure*}
${\sf tanh}$ is a non-linear activation function that converts any given value to a value
between -1 and 1, and ${\sf softmax}$ converts a given $K$-sized vector of arbitrary values to
another $K$-sized vector of values in the range $[0,1]$ that sum to 1---essentially a probability
distribution.

The type of edge at time $i$ decides which RNN to choose to update the (shared) hidden state $h_i$
and the output $y_i$. Training consists of learning values for the entries in all the ${\bf W}$ and
${\bf b}$ matrices.  During training, $v_i'$, $e_i$ and the target output are known from the data
point, and so we optimize a standard cross-entropy loss function (over all $i$) between the output
$y_i$ and the target output. During inference, $P(v_{i+1} | {\bf Y}_i, \Sz)$ is simply the
probability distribution $y_i$, the result of the ${\sf softmax}$.

A sketch is obtained by starting with the root node pair $(v_1, e_1) = (\rootnode, child)$,
recursively applying Equation~\ref{eqn:rnnoutput} to get the output distribution $y_i$, sampling a
value for $v_{i+1}$ from $y_i$, and growing the tree by adding the sampled node to it. The edge
$e_{i+1}$ is provided as $c$ or $s$ depending on the $v_{i+1}$ that was sampled. If only one type of
edge is feasible (for instance, if the node is a terminal in the grammar, only a $sibling$ edge is
possible with the next node), then only that edge is provided. If both edges are feasible, then both
possibilities are recursively explored, growing the tree in both directions.

\paragraph{Remarks.} In our implementation, we generate trees in a depth-first fashion, by exploring
a $child$ edge before a $sibling$ edge if both are possible. If a node has two children, a neural
encoding of the nodes that were generated on the left is carried onto the right sub-tree so that the
generation of this tree can leverage additional information about its previously generated sibling.
We refer the reader to Section 2.4 of~\citet{treeRNN} for more details.

%% file: eval-appendix.tex
In this section, we provide results of additional experimental evaluation. 

\subsection{Clustering}

\begin{figure}
\begin{center}
\includegraphics[scale=0.4]{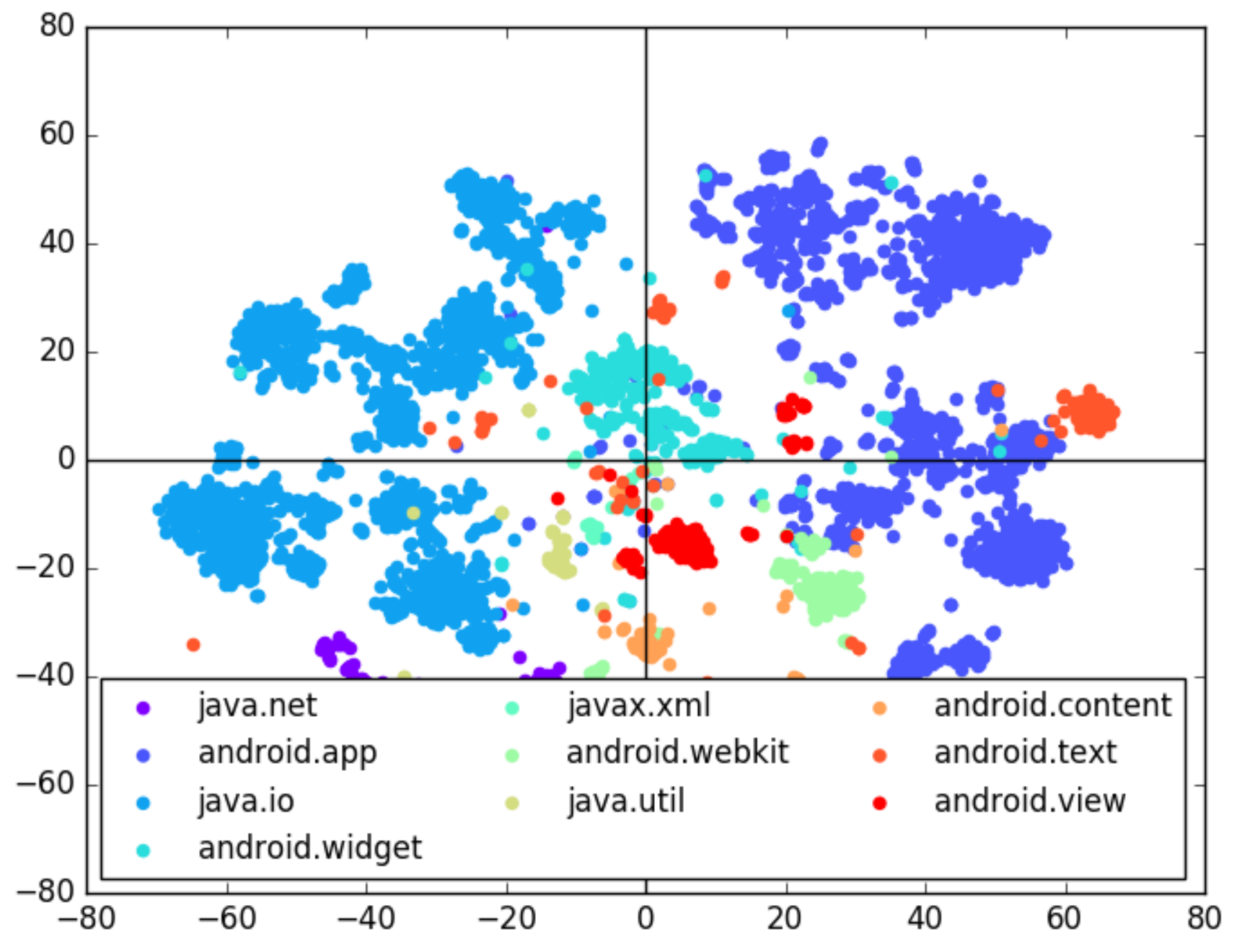}
\end{center}
\caption{2-dimensional projection of latent space of the \GSNN-Sk model}
\figlabel{2dplot-GSNN}
\end{figure}

Similar to the visualization of the 2-dimensional latent space in
\figref{2dplot}, we also plotted the latent space of the \GSNN-Sk model trained on
sketches. \figref{2dplot-GSNN} shows this plot. We observed that the latent
space is clustered, relatively, more densely than that of our model (keep in
mind that the plot colors are different when comparing them).


\subsection{Qualitative Evaluation}

\begin{figure*}
  \begin{tabular}{|cc|} \hline
    {\bf Input to \system} & {\bf Generated program ranked among the top-5} \\ \hline
    \begin{minipage}{0.3\textwidth}
      \begin{tabular}{l}
      $\Sx_{\Types} = \{\texttt{FileWriter}\}$ \\
      $\Sx_{\Calls} = \{\texttt{write}\}$ \\
      $\Sx_{\Kws} = \emptyset$
      \end{tabular}
    \end{minipage}
    &
    \begin{minipage}{0.65\textwidth}
      \begin{lstlisting}
        BufferedWriter bw;
        FileWriter fw;
        try {
            fw = new FileWriter($String, $boolean);
            bw = new BufferedWriter(fw);
            bw.write($String);
            bw.newLine();
            bw.flush();
            bw.close();
        } catch (IOException _e) {
        }
      \end{lstlisting}
    \end{minipage}
    \\ \hline

    \begin{minipage}{0.3\textwidth}
      \begin{tabular}{l}
      $\Sx_{\Types} = \emptyset$ \\
      $\Sx_{\Calls} = \emptyset$ \\
      $\Sx_{\Kws} = \{$android, dialog, set, \\ ~~~~title, message$\}$
      \end{tabular}
    \end{minipage}
    &
    \begin{minipage}{0.65\textwidth}
      \begin{lstlisting}
        Builder builder2;
        Builder builder1;
        AlertDialog alertDialog;
        Builder builder4;
        Builder builder3;
        builder1 = new Builder($Context);
        builder2 = builder1.setTitle($String);
        builder3 = builder2.setMessage($String);
        builder4 = builder3.setNeutralButton($String,
                                $OnClickListener);
        alertDialog = builder4.show();
      \end{lstlisting}
    \end{minipage}
    \\ \hline

    \begin{minipage}{0.3\textwidth}
      \begin{tabular}{l}
      $\Sx_{\Types} = \emptyset$ \\
      $\Sx_{\Calls} = \{\small{\texttt{startPreview}}\}$ \\
      $\Sx_{\Kws} = \emptyset$
      \end{tabular}
    \end{minipage}
    &
    \begin{minipage}{0.7\textwidth}
      \begin{lstlisting}
        Parameters parameters;
        parameters = $Camera.getParameters();
        parameters.setPreviewSize($int, $int);
        parameters.setRecordingHint($boolean);
        $Camera.setParameters(parameters);
        $Camera.startPreview();
      \end{lstlisting}
    \end{minipage}
    \\ \hline
  \end{tabular}
  \caption{Qualitative usage scenarios of \system.}
  \figlabel{qualitative}
\end{figure*}

To give a sense of the quality of the end-to-end generation, we present and
discuss a few usage scenarios for our system, \system.  In each scenario, we
started with a set of API calls, types or keywords as labels that indicate what
we (as the user) would like the generated code to perform.  We then pick a
single program in the top-5 results returned by \system and discuss it.
\figref{qualitative} shows three such example usage scenarios.

In the first scenario, we would like the system to generate a program to write
something to a file by calling ${\tt write}$ using the type ${\tt FileWriter}$.
With this label, we invoked \system and it returned with a program that actually
accomplishes the task. Note that even though we only specified ${\tt
FileWriter}$, the program uses it to feed a ${\tt BufferedWriter}$ to write to a
file. This is an interesting pattern learned from data, that file reads and writes
in Java often take place in a buffered manner. Also note that the program
correctly flushes the buffer before closing it, even though none of this was
explicitly specified in the input.

In the second scenario, we would like the generated program to set the title and
message of an Android dialog. This time we provide no API calls or types but
only keywords. With this, \system generated a program that first builds an
Android dialog box using the helper class ${\tt AlertDialog.Builder}$, and does
set its title and message. In addition, the program also adds a button to the
dialog box -- another interesting pattern learned from data that dialog boxes in
Android often have a button, typically to close the dialog. Finally it shows the
dialog with these items.

In the final scenario, we would like \system to generate code to start preview
mode in the phone's camera. We provided simply the API call ${\tt startPreview}$
as input. With this, the system was automatically able to recognize that we are
interested in the camera API, and generate a program that accomplishes the task.
Note that the program first obtains the camera parameters, and sets the preview
display size (the ${\tt int}$ arguments are the width and height) before
starting the preview. We confirmed from the Android ${\tt Camera}$ API
documentation that this is recommended practice, and the model appears to
have learned this automatically from data.

%% file: ms.bbl
\begin{thebibliography}{37}
\providecommand{\natexlab}[1]{#1}
\providecommand{\url}[1]{\texttt{#1}}
\expandafter\ifx\csname urlstyle\endcsname\relax
  \providecommand{\doi}[1]{doi: #1}\else
  \providecommand{\doi}{doi: \begingroup \urlstyle{rm}\Url}\fi

\bibitem[and(2017)]{androiddrawer}
Androiddrawer.
\newblock \url{http://www.androiddrawer.com}, 2017.

\bibitem[Abadi et~al.(2015)Abadi, Agarwal, Barham, Brevdo, Chen, Citro,
  Corrado, Davis, Dean, Devin, Ghemawat, Goodfellow, Harp, Irving, Isard, Jia,
  Jozefowicz, Kaiser, Kudlur, Levenberg, Mane, Monga, Moore, Murray, Olah,
  Schuster, Shlens, Steiner, Sutskever, Talwar, Tucker, Vanhoucke, Vasudevan,
  Viegas, Vinyals, Warden, Wattenberg, Wicke, Yu, and Zheng]{tensorflow}
Martin Abadi, Ashish Agarwal, Paul Barham, Eugene Brevdo, Zhifeng Chen, Craig
  Citro, Greg Corrado, Andy Davis, Jeffrey Dean, Matthieu Devin, Sanjay
  Ghemawat, Ian Goodfellow, Andrew Harp, Geoffrey Irving, Michael Isard,
  Yangqing Jia, Rafal Jozefowicz, Lukasz Kaiser, Manjunath Kudlur, Josh
  Levenberg, Dan Mane, Rajat Monga, Sherry Moore, Derek Murray, Chris Olah,
  Mike Schuster, Jonathon Shlens, Benoit Steiner, Ilya Sutskever, Kunal Talwar,
  Paul Tucker, Vincent Vanhoucke, Vijay Vasudevan, Fernanda Viegas, Oriol
  Vinyals, Pete Warden, Martin Wattenberg, Martin Wicke, Yuan Yu, and Xiaoqiang
  Zheng.
\newblock Tensorflow: Large-scale machine learning on heterogeneous distributed
  systems, 2015.
\newblock URL \url{http://download.tensorflow.org/paper/whitepaper2015.pdf}.

\bibitem[Allamanis \& Sutton(2014)Allamanis and Sutton]{idiomsFSE14}
Miltiadis Allamanis and Charles Sutton.
\newblock Mining idioms from source code.
\newblock In \emph{FSE}, pp.\  472--483, 2014.

\bibitem[Alur et~al.(2013)Alur, Bod\'{\i}k, Juniwal, Martin, Raghothaman,
  Seshia, Singh, Solar-Lezama, Torlak, and Udupa]{sygus}
Rajeev Alur, Rastislav Bod\'{\i}k, Garvit Juniwal, Milo M.~K. Martin, Mukund
  Raghothaman, Sanjit~A. Seshia, Rishabh Singh, Armando Solar-Lezama, Emina
  Torlak, and Abhishek Udupa.
\newblock Syntax-guided synthesis.
\newblock In \emph{FMCAD}, pp.\  1--17, 2013.

\bibitem[Balog et~al.(2017)Balog, Gaunt, Brockschmidt, Nowozin, and
  Tarlow]{balog2016deepcoder}
Matej Balog, Alexander~L Gaunt, Marc Brockschmidt, Sebastian Nowozin, and
  Daniel Tarlow.
\newblock Deepcoder: Learning to write programs.
\newblock In \emph{ICLR}, 2017.

\bibitem[Bielik et~al.(2016)Bielik, Raychev, and Vechev]{bielik2016phog}
Pavol Bielik, Veselin Raychev, and Martin~T Vechev.
\newblock {PHOG}: probabilistic model for code.
\newblock In \emph{ICML}, pp.\  19--24, 2016.

\bibitem[Devlin et~al.(2017)Devlin, Uesato, Bhupatiraju, Singh, Mohamed, and
  Kohli]{robustfill}
Jacob Devlin, Jonathan Uesato, Surya Bhupatiraju, Rishabh Singh, Abdel{-}rahman
  Mohamed, and Pushmeet Kohli.
\newblock Robustfill: Neural program learning under noisy {I/O}.
\newblock In \emph{ICML}, 2017.

\bibitem[Feser et~al.(2015)Feser, Chaudhuri, and Dillig]{example-pldi15}
John~K. Feser, Swarat Chaudhuri, and Isil Dillig.
\newblock Synthesizing data structure transformations from input-output
  examples.
\newblock In \emph{PLDI}, pp.\  229--239. ACM, 2015.

\bibitem[Gaunt et~al.(2016)Gaunt, Brockschmidt, Singh, Kushman, Kohli, Taylor,
  and Tarlow]{gaunt2016terpret}
Alexander~L Gaunt, Marc Brockschmidt, Rishabh Singh, Nate Kushman, Pushmeet
  Kohli, Jonathan Taylor, and Daniel Tarlow.
\newblock Terpret: A probabilistic programming language for program induction.
\newblock \emph{arXiv preprint arXiv:1608.04428}, 2016.

\bibitem[Gulwani(2011)]{example-popl11}
Sumit Gulwani.
\newblock Automating string processing in spreadsheets using input-output
  examples.
\newblock In \emph{POPL}, pp.\  317--330. ACM, 2011.

\bibitem[Gulwani et~al.(2015)Gulwani, Hern\'{a}ndez-Orallo, Kitzelmann,
  Muggleton, Schmid, and Zorn]{inductive-realworld}
Sumit Gulwani, Jos{\'e} Hern\'{a}ndez-Orallo, Emanuel Kitzelmann, Stephen~H.
  Muggleton, Ute Schmid, and Benjamin Zorn.
\newblock Inductive programming meets the real world.
\newblock \emph{Communications of the ACM}, 58\penalty0 (11):\penalty0 90--99,
  2015.

\bibitem[Gulwani et~al.(2017)Gulwani, Polozov, and Singh]{gulwani-survey}
Sumit Gulwani, Oleksandr Polozov, and Rishabh Singh.
\newblock Program synthesis.
\newblock \emph{Foundations and Trends in Programming Languages}, 4\penalty0
  (1-2):\penalty0 1--119, 2017.

\bibitem[Ha \& Eck(2017)Ha and Eck]{ha2017neural}
David Ha and Douglas Eck.
\newblock A neural representation of sketch drawings.
\newblock \emph{arXiv preprint arXiv:1704.03477}, 2017.

\bibitem[Hindle et~al.(2012)Hindle, Barr, Su, Gabel, and
  Devanbu]{hindle2012naturalness}
Abram Hindle, Earl~T Barr, Zhendong Su, Mark Gabel, and Premkumar Devanbu.
\newblock On the naturalness of software.
\newblock In \emph{ICSE}, pp.\  837--847, 2012.

\bibitem[Kingma \& Ba(2014)Kingma and Ba]{kingma2014adam}
Diederik Kingma and Jimmy Ba.
\newblock Adam: A method for stochastic optimization.
\newblock \emph{arXiv preprint arXiv:1412.6980}, 2014.

\bibitem[Kingma \& Welling(2014)Kingma and Welling]{kingma2013auto}
Diederik~P Kingma and Max Welling.
\newblock Auto-encoding variational bayes.
\newblock In \emph{ICLR}, 2014.

\bibitem[Kusner et~al.(2017)Kusner, Paige, and
  Hern{\'a}ndez-Lobato]{kusner2017grammar}
Matt~J Kusner, Brooks Paige, and Jos{\'e}~Miguel Hern{\'a}ndez-Lobato.
\newblock Grammar variational autoencoder.
\newblock \emph{arXiv preprint arXiv:1703.01925}, 2017.

\bibitem[Ling et~al.(2016)Ling, Grefenstette, Hermann, Ko{\v{c}}isk{\`y},
  Senior, Wang, and Blunsom]{ling2016latent}
Wang Ling, Edward Grefenstette, Karl~Moritz Hermann, Tom{\'a}{\v{s}}
  Ko{\v{c}}isk{\`y}, Andrew Senior, Fumin Wang, and Phil Blunsom.
\newblock Latent predictor networks for code generation.
\newblock \emph{arXiv preprint arXiv:1603.06744}, 2016.

\bibitem[Maaten \& Hinton(2008)Maaten and Hinton]{maaten2008tsne}
Laurens van~der Maaten and Geoffrey Hinton.
\newblock Visualizing data using t-sne.
\newblock \emph{Journal of Machine Learning Research}, 9\penalty0
  (Nov):\penalty0 2579--2605, 2008.

\bibitem[Maddison \& Tarlow(2014)Maddison and Tarlow]{ICML:MT14}
C.J. Maddison and D.~Tarlow.
\newblock Structured generative models of natural source code.
\newblock In \emph{ICML}, 2014.

\bibitem[Manna \& Waldinger(1971)Manna and Waldinger]{mannawaldinger}
Zohar Manna and Richard~J. Waldinger.
\newblock Toward automatic program synthesis.
\newblock \emph{Communications of the {ACM}}, 14\penalty0 (3):\penalty0
  151--165, 1971.

\bibitem[Nguyen \& Nguyen(2015)Nguyen and Nguyen]{nguyenICSE15}
Anh~Tuan Nguyen and Tien~N. Nguyen.
\newblock Graph-based statistical language model for code.
\newblock In \emph{ICSE}, pp.\  858--868, 2015.

\bibitem[Nguyen et~al.(2013)Nguyen, Nguyen, Nguyen, and Nguyen]{nguyenFSE13}
Tung~Thanh Nguyen, Anh~Tuan Nguyen, Hoan~Anh Nguyen, and Tien~N. Nguyen.
\newblock A statistical semantic language model for source code.
\newblock In \emph{ESEC/FSE}, pp.\  532--542, 2013.

\bibitem[Oord et~al.(2016)Oord, Kalchbrenner, and Kavukcuoglu]{oord2016pixel}
Aaron van~den Oord, Nal Kalchbrenner, and Koray Kavukcuoglu.
\newblock Pixel recurrent neural networks.
\newblock \emph{arXiv preprint arXiv:1601.06759}, 2016.

\bibitem[Parisotto et~al.(2016)Parisotto, Mohamed, Singh, Li, Zhou, and
  Kohli]{parisotto2016neuro}
Emilio Parisotto, Abdel-rahman Mohamed, Rishabh Singh, Lihong Li, Dengyong
  Zhou, and Pushmeet Kohli.
\newblock Neuro-symbolic program synthesis.
\newblock \emph{arXiv preprint arXiv:1611.01855}, 2016.

\bibitem[Polozov \& Gulwani(2015)Polozov and Gulwani]{flashmeta}
Oleksandr Polozov and Sumit Gulwani.
\newblock Flashmeta: A framework for inductive program synthesis.
\newblock In \emph{OOPSLA}, volume~50, pp.\  107--126, 2015.

\bibitem[Rabinovich et~al.(2017)Rabinovich, Stern, and
  Klein]{rabinovich2017abstract}
Maxim Rabinovich, Mitchell Stern, and Dan Klein.
\newblock Abstract syntax networks for code generation and semantic parsing.
\newblock \emph{arXiv preprint arXiv:1704.07535}, 2017.

\bibitem[Raychev et~al.(2014)Raychev, Vechev, and Yahav]{raychev2014code}
Veselin Raychev, Martin Vechev, and Eran Yahav.
\newblock Code completion with statistical language models.
\newblock In \emph{PLDI}, 2014.

\bibitem[Riedel et~al.(2016)Riedel, Bosnjak, and
  Rockt{\"a}schel]{riedel2016programming}
Sebastian Riedel, Matko Bosnjak, and Tim Rockt{\"a}schel.
\newblock Programming with a differentiable forth interpreter.
\newblock \emph{CoRR, abs/1605.06640}, 2016.

\bibitem[Schkufza et~al.(2016)Schkufza, Sharma, and Aiken]{Schkufza0A16}
Eric Schkufza, Rahul Sharma, and Alex Aiken.
\newblock Stochastic program optimization.
\newblock \emph{Commun. {ACM}}, 59\penalty0 (2):\penalty0 114--122, 2016.

\bibitem[Skylot(2017)]{JADX}
Skylot.
\newblock {JADX}: {D}ex to {J}ava decompiler.
\newblock \url{https://github.com/skylot/jadx}, 2017.

\bibitem[Sohn et~al.(2015)Sohn, Lee, and Yan]{GSNN}
Kihyuk Sohn, Honglak Lee, and Xinchen Yan.
\newblock Learning structured output representation using deep conditional
  generative models.
\newblock In \emph{NIPS}, pp.\  3483--3491, 2015.

\bibitem[Solar-Lezama et~al.(2006)Solar-Lezama, Tancau, Bod\'{\i}k, Seshia, and
  Saraswat]{armandosketching}
Armando Solar-Lezama, Liviu Tancau, Rastislav Bod\'{\i}k, Sanjit~A. Seshia, and
  Vijay~A. Saraswat.
\newblock Combinatorial sketching for finite programs.
\newblock In \emph{ASPLOS}, pp.\  404--415, 2006.

\bibitem[Summers(1977)]{summers1977methodology}
Phillip~D Summers.
\newblock A methodology for {LISP} program construction from examples.
\newblock \emph{Journal of the ACM (JACM)}, 24\penalty0 (1):\penalty0 161--175,
  1977.

\bibitem[Vinyals et~al.(2015)Vinyals, Toshev, Bengio, and
  Erhan]{vinyals2015show}
Oriol Vinyals, Alexander Toshev, Samy Bengio, and Dumitru Erhan.
\newblock Show and tell: A neural image caption generator.
\newblock In \emph{CVPR}, pp.\  3156--3164, 2015.

\bibitem[Yin \& Neubig(2017)Yin and Neubig]{yin2017syntactic}
Pengcheng Yin and Graham Neubig.
\newblock A syntactic neural model for general-purpose code generation.
\newblock \emph{arXiv preprint arXiv:1704.01696}, 2017.

\bibitem[Zhang et~al.(2016)Zhang, Lu, and Lapata]{treeRNN}
Xingxing Zhang, Liang Lu, and Mirella Lapata.
\newblock Top-down tree long short-term memory networks.
\newblock In \emph{{NAACL}-{HLT}}, pp.\  310--320, 2016.

\end{thebibliography}
